\documentclass[journal]{IEEEtran}
\IEEEoverridecommandlockouts
\usepackage{algorithmic}
\usepackage{algorithm}
\usepackage{amsmath,amssymb,amsfonts}
\usepackage{array}
\usepackage{booktabs}
\usepackage{cite}
\usepackage{comment}
\usepackage{csquotes}
\usepackage{epsfig}
\usepackage{graphicx}
\usepackage{siunitx}
\usepackage{stfloats}
\usepackage{tabularx}
\usepackage{textcomp}
\usepackage{xcolor}
\usepackage{upgreek}
\usepackage{url}
\usepackage{verbatim}
\usepackage{orcidlink}
\usepackage[absolute]{textpos}
\textblockorigin{0in}{0in} 
\def\BibTeX{{\rm B\kern-.05em{\sc i\kern-.025em b}\kern-.08em T\kern-.1667em\lower.7ex\hbox{E}\kern-.125emX}}
\DeclareSIUnit\bar{bar}
\begin{document}
\title{Growth and microwave properties of FeSe thin films and comparison with Fe(Se,Te)}
\author{
Alessandro Magalotti~\orcidlink{0009-0004-3352-1977},~\IEEEmembership{Student Member,~IEEE},
Andrea Alimenti~\orcidlink{0000-0002-4459-6147},~\IEEEmembership{Member,~IEEE},
Valeria Braccini~\orcidlink{0000-0003-0073-367X},
Giuseppe Celentano~\orcidlink{0000-0001-6017-0739},
Matteo Cialone~\orcidlink{0000-0002-3018-2787},
Antonella Mancini~\orcidlink{0000-0002-4148-1010},
Andrea Masi~\orcidlink{0000-0002-1976-0603},
Nicola Pompeo~\orcidlink{0000-0003-4847-1234},~\IEEEmembership{Senior~Member,~IEEE},
Enrico Silva~\orcidlink{0000-0001-8633-4295},~\IEEEmembership{Senior~Member,~IEEE},
Giovanni Sotgiu~\orcidlink{0000-0003-2841-9316},
Kostiantyn Torokhtii~\orcidlink{0000-0002-3420-3864},~\IEEEmembership{Member,~IEEE},
Pablo Vidal Garc\'ia~\orcidlink{0000-0003-4847-1234},~\IEEEmembership{Member,~IEEE},
Angelo Vannozzi~\orcidlink{0000-0003-4628-4312}
\thanks{Received 13 October 2025; revised 15 December 2025; accepted 4 February 2026. Date of publication 16 February 2026; date of current version 24 February 2026. This work was supported in part by MIUR-PRIN Project “HIBiSCUS” under Grant 201785KWLE and in part by IronMOON under Grant 2022BPJL2L.(Corresponding author: Alessandro Magalotti.)}
\thanks{Alessandro Magalotti, Andrea Alimenti, Nicola Pompeo, Enrico Silva, Giovanni Sotgiu, Kostiantyn Torokhtii, and Pablo Vidal García are with the Department of Industrial, Electronic and Mechanical Engineering, Roma Tre University, 00154 Rome, Italy (e-mail: alessandro.magalotti@uniroma3.it).}
\thanks{Valeria Braccini is with SPIN-CNR, 16152 Genoa, Italy.}
\thanks{Giuseppe Celentano, Antonella Mancini, Andrea Masi, and Angelo Vannozzi are with ENEA, 00044 Frascati, Italy (e-mail: angelo.vannozzi@enea.it).}
\thanks{Matteo Cialone is with the Department of Physics, University of Genoa, 16152 Genoa, Italy, and also with SPIN-CNR, 16152 Genoa, Italy.}
\thanks{Color versions of one or more figures in this article are available at https://doi.org/10.1109/TASC.2026.3665171.
}
\thanks{Digital Object Identifier 10.1109/TASC.2026.3665171}
}
\markboth{IEEE TRANSACTIONS ON APPLIED SUPERCONDUCTIVITY, VOL. 36, NO. 5, AUGUST 2026}%
{}
\maketitle
\begin{textblock}{7}(0.75,10.5) 
\noindent\centering
\copyright~2026 The Authors. This work is licensed under a Creative Commons Attribution 4.0 License. For more information, see https://creativecommons.org/licenses/by/4.0/
\end{textblock}
\begin{abstract}
In this work, we have grown $\sim$\qty{100}{\nano\meter} thick pristine FeSe films by pulsed laser deposition. The films were structurally characterized with X-ray diffraction and their surface morphology checked through atomic force microscopy. Microwave measurements, performed with a dielectric loaded resonator tuned at the frequency of \qty{8}{\giga\hertz}, allowed the characterization of the samples surface resistance, in view of potential applications in microwave haloscopes for dark matter search. Here, we report the comparison of the microwave properties of FeSe with Fe(Se,Te) thin films, as the temperature is swept from \qtyrange{4}{20}{\kelvin}. By applying a constant static magnetic field of \qty{12}{\tesla}, it was also possible to discern the magnetic field resilience of the two samples. FeSe showed a larger critical temperature drift as the field is applied and a small broadening, while the opposite appears in Fe(Se,Te). A preliminary analysis of vortex pinning shows margins for optimizing pinning in FeSe.
\end{abstract}
\section{Introduction}
\label{sec:intro}
\IEEEPARstart{I}{n} the hunt for intergalactic dark matter \cite{kim2021cosmic}, huge efforts are currently dedicated in the design and fabrication of reliable detectors. Indeed, given the unknown mass of the axion, $i.e.$ the theoretically predicted particle that should prove and justify the existence of cold dark matter \cite{Peccei1977,Sikivie1985}, a large energy spectrum needs to be scanned. At microwaves, one of the proposed devices is the haloscope \cite{semertzidis2022axion}, a microwave resonant cavity that needs to be exposed to strong static magnetic fields of the order of several \unit{\tesla}. Due to the Primakov effect \cite{Inverse_primakoff}, the axion can interact with the magnetic field and be converted into a photon, that can be revealed with the resonant cavity. In order to enhance the cavity sensitivity and scan rate, materials with extremely high conductivity are necessary to internally coat the cavity walls in order to obtain high cavity quality factors \cite{Posen_2023}. Since the cavities will in any case operate at cryogenic temperatures (${T\ll\qty{4.2}{\kelvin}}$) to reduce the thermal noise \cite{Cervantes_2024}, superconductors are an immediate choice. However, their resilience in strong dc magnetic fields must be demonstrated on a case-by-case basis: in fact, vortex motion dissipation overcomes by orders of magnitude the residual dissipation due to impurities or quasiparticles in zero magnetic field \cite{bardeen1965theory}, and pinning properties acting effectively at microwave frequencies are the most relevant material engineering issue \cite{gurevich2014challenges}. Among all the families, iron-based superconductors (IBS) have the potential to be great candidates: on one side, they have strong upper critical fields \cite{hosono2018recent}, with presumably low microwave dissipation in high fields; in addition, IBS can in principle be electrochemically deposited \cite{Piperno_2023,Piperno_2024}, possibly on arbitrary shapes, a significant, potential advantage with respect to YBCO tapes that need to be glued to the cavity walls \cite{ahn_2021_prototype}. On the other hand, IBS performance are strongly dependent on intrinsic properties, like the chemical composition or the defects present in the lattice, but are also affected by external parameters, like the material thickness or the substrate on which are grown \cite{hosono2015a}.
Encouraged by previous results (see \cite{Alimenti_2022}) and in order to further explore some of the IBS family compounds, in this work we have successfully grown FeSe thin films by Pulsed Laser Deposition (PLD). Its microwave properties were compared to a previously studied Fe(Se,Te) pristine thin films \cite{Pompeo_2020_Sust,Magalotti_2025_Arxiv}. Microwave measurement were taken at frequencies $\sim$\qty{10}{\giga\hertz}, in a strong dc magnetic field ${\mu_0H=\qty{12}{\tesla}}$ and for temperatures in the range \qtyrange{4}{20}{\kelvin}.
\section{Samples}
\label{sec:samples}
\begin{figure}
    \centering
    \includegraphics[width=0.8\linewidth]{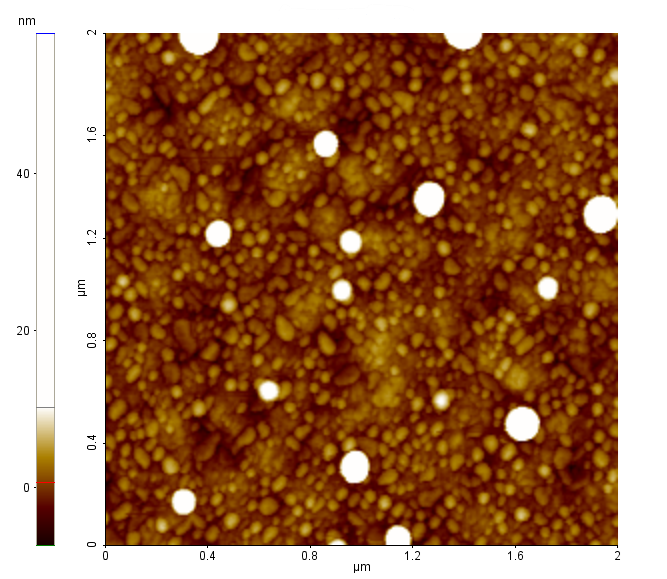}
    \caption{AFM image of the FeSe film surface over a $2\,\unit{\micro\meter}\times2\,\unit{\micro\meter}$ area.}
    \label{fig:AFM}
\end{figure}

\begin{figure}
    \centering
    \includegraphics[width=0.9\linewidth]{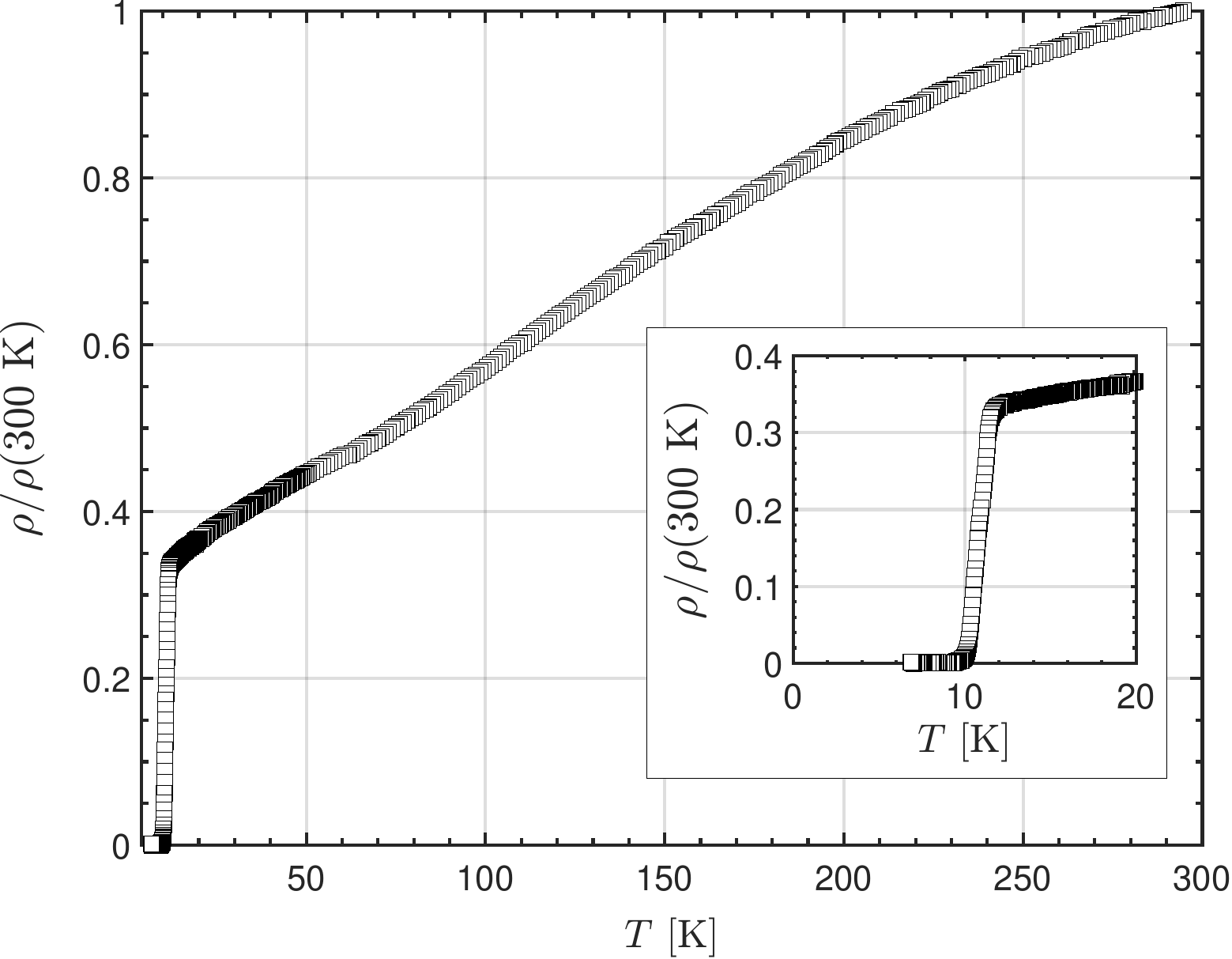}
    \caption{Normalized electrical resistivity for the FeSe film as a function of temperature, with respect to the room temperature value. The inset highlights the superconducting transition region.}
    \label{fig:DC_Tsw}
\end{figure}

Thin FeSe films were synthesized by PLD using the fourth harmonics of a Nd:YAG laser (${\lambda_4=\qty{266}{\nano\meter}}$) with the laser fluence of \qty{1}{\joule\cdot\centi\metre^{-2}}, a repetition rate of \qty{3}{\hertz}, and \qty{45}{\milli\meter} target-to-substrate distance. The target was prepared by solid-state reaction from high-purity Fe and Se powders \cite{Masi_2020}. Depositions were performed on ${\text{(00$l$)-oriented}}$ CaF$_2$ substrates at the temperature ${T_s = \qty{300}{\celsius}}$ under high-vacuum (${\text{pressure}\sim\SI{E-6}{\milli\bar}}$), followed by in-vacuum cooling. The deposition time was adjusted to achieve a film thickness of about ${\sim\qty{100}{\nano\meter}}$.
Structural characterization was performed by X-ray diffraction (XRD) in $\theta$-$2\theta$ geometry and by ${\omega\text{-scan}}$. Surface morphology was examined using a Park Systems XE-150 Atomic Force Microscope (AFM) operating in non-contact mode at room temperature and under ambient atmosphere. Pre-mounted non-contact, high-resolution cantilevers working at \qty{309}{\mega\hertz} with nominal tip radius below \qty{10}{\nano\meter} were used. Images were flattened by subtracting a linear background in the fast scan direction and a quadratic background in the slow scan direction. The temperature dependence of resistivity was measured using the four-probe method. The critical temperature $T_{C0}$ was evaluated as zero-resistance temperature, and $T_{Con}$ corresponds to $90\%$ of the normal state resistivity. 
\begin{figure}[h]
    \centering
    \includegraphics[width=0.8\linewidth]{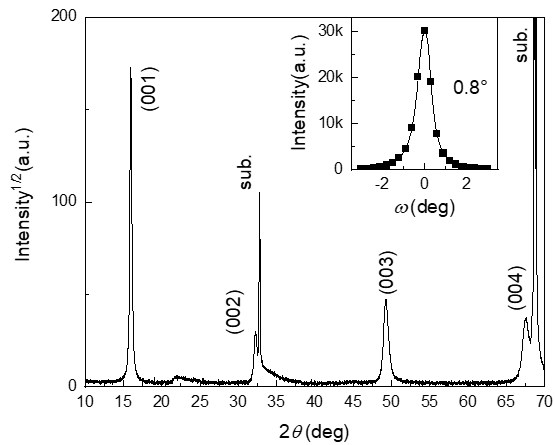}
    \caption{$\theta$-$2\theta$ XRD pattern of the FeSe film on CaF$_2$ substrate. In the inset the rocking curve of the (001) reflection is shown.}
    \label{fig:XDR}
\end{figure}
The $\theta$-$2\theta$ X-ray diffraction pattern of the FeSe film, shown in \text{Figure \ref{fig:XDR}}, displays distinct peaks attributable to both the substrate and the tetragonal FeSe phase. The exclusive presence of (00$l$) FeSe reflections indicates a preferred \text{out-of-plane} alignment of the film along the c-axis. The narrow \text{full-width} at half maximum (FWHM) of ${\sim\qty{0.8}{\degree}}$, observed in the rocking curve (inset of Figure~\ref{fig:XDR}), suggests a sharp out-of-plane orientation distribution. The c-axis parameter (${\approx\qty{0.555}{\nano\meter}}$) is consistent with the lattice parameters reported for FeSe films on CaF$_2$ with comparable thickness \cite{nabeshima2013enhancement,qiu2017interface}.
In Figure \ref{fig:AFM}, the AFM image of the FeSe film surface is reported. The surface appears smooth and continuous, with no visible cracks or voids. The morphology of FeSe film includes fine equiaxed grains with average lateral size of \qty{42}{\nano\meter} \text{(SD 11)} and average height of \qty{4}{\nano\meter} (SD 2). A few platelet-like grains are sometimes visible (${\text{density}\sim\qty{1}{\micro\meter^{-2}}}$), as high as the grains and approximately \qty{150}{\nano\meter} long, often oriented along the ${<100>}$ film direction. Large, rounded droplets appear on the surface, with an average diameter of \qty{93}{\nano\meter} (SD 23) and height \qty{34}{\nano\meter} (SD 16). Film roughness is \qty{5.8}{\nano\meter} (rms), decreasing to \qty{2.1}{\nano\meter} if droplets are excluded. A \text{hill-and-valley} film background modulation is visible, with hills and valleys \qtyrange{200}{250}{\nano\meter} wide and a few \unit{\nano\meter} above or below the average film level.
The temperature dependence of the electrical resistance for the FeSe film is shown in Figure \ref{fig:DC_Tsw}. The film exhibits a superconducting transition with an onset critical temperature ${T_{Con}\simeq\qty{12}{\kelvin}}$ and a zero-resistance temperature ${T_{C0}\simeq\qty{10}{\kelvin}}$. These values exceed those typically reported for bulk FeSe, which generally shows $T_{Con}$ values in the range \qtyrange{7}{8}{\kelvin}. The enhanced critical temperatures observed in thin films are consistent with previous studies attributing such improvements to substrate-induced strain, interface effects, and confirm the good crystalline quality of our films \cite{feng2018tunable}.
Samples of nominal composition FeSe$_{0.5}$Te$_{0.5}$ (shortly named Fe(Se,Te)) were deposited similarly to FeSe thin films, by PLD on CaF$_2$ substrates, to reduce differences when comparing the two IBS. The growth and characterization of such samples was extensively discussed previously \cite{Iebole_2024}. Microwave properties of samples similar to the one here investigated have been reported previously in moderate magnetic fields ($\mu_0H \leq \qty{1.2}{\tesla}$)\cite{Pompeo_2020_Sust,Pompeo_2021_TAS}. Table \ref{tab:T1} reports the main geometrical parameters of the two IBS films investigated. 
\begin{table}
    \centering
    \renewcommand{\arraystretch}{1.3}
    \begin{tabular}{|c|c|c|c|c|c|}
    \hline
    Sample & Thickness & Substrate & Size \\
    \hline    
    Fe(Se,Te) & $275\,\unit{\nano\meter}$ & CaF$_2$ & $10\times10\,\unit{\milli\meter^2}$ \\
    \hline
    FeSe & $80\,\unit{\nano\meter}$ & CaF$_2$ & $7\times7\,\unit{\milli\meter^2}$ \\
    \hline
    \end{tabular}
    \vspace{5 pt}
    \caption{IBS samples properties}
    \label{tab:T1}
\end{table}
\section{Microwave measurements: method}
\label{sec:method}
At microwave frequencies the electrical response function of a material is the complex surface impedance, ${Z_s = R_s + \text{i}X_s}$, where $R_s$ and $X_s$ are the surface resistance and surface reactance, respectively \cite{chen2004microwave}. $Z_s$ in good conductors and superconductors depends on the complex resistivity and on geometrical effects arising from partial penetration of the electromagnetic field across the (super)conducting film. In the particular case of a thin conducting film deposited onto a dielectric substrate, one can approximate  \cite{Pompeo2025FiniteThickness}:
\begin{equation}
    Z_s \simeq \frac{\tilde{\rho}}{d},
    \label{eq:thin}
\end{equation}
where $\tilde{\rho}$ is the complex resistivity and $d$ the film thickness. For a superconductor, the approximation is valid when the penetration depth ${\lambda < d}$. Given the samples thicknesses reported in Table \ref{tab:T1},  Equation \ref{eq:thin} can be safely applied.
In order to measure $R_s$ of the IBS samples, a cylindrical dielectrically loaded copper resonator, with radius of $\qty{12}{\milli\metre}$, was employed, as the one described in \cite{alimenti2019challenging}. Rutile (TiO$_2$) and sapphire (Al$_2$O$_3$) dielectric cylindrical pucks were used to yield resonant frequencies of ${\sim\qty{8}{\giga\hertz}}$ and ${\sim\qty{15}{\giga\hertz}}$ for measurements in FeSe and Fe(Se,Te), respectively \cite{alimenti2019challenging}. By inserting the sample between the optically polished crystal and one of the resonator copper bases, differential measurements in the so-called end-wall perturbation technique were performed\cite{alimenti2019challenging}. The transverse electric mode TE$_{011}$ was excited in the resonator, with circular electric currents induced on the sample surface. The microwave device was inserted into a superconducting cryomagnet, capable to generate a maximum magnetic field $\mu_0H=\qty{12}{\tesla}$ in its bore, applied perpendicular to the sample surface. The measurements of the scattering matrix $S_{ij}$ of the two port resonator with a vector network analyzer as a function of the temperature at fixed magnetic field yielded the quality factor $Q$, whence the surface resistance of the sample under study by means of the relation \cite{chen2004microwave,alimenti2019challenging}:
\begin{equation}
    \Delta R_s = R_s - R_{s,0} = G\left(\frac{1}{Q}-\frac{1}{Q_0}\right)
\end{equation}
where $R_{s,0}$ and $Q_0$ are the reference values measured at ${T^\star = \min(T)}$ and $\mu_0H=0$, and $G$ is the sample geometrical factor, obtained through electromagnetic simulations.
\section{Microwave measurements: results}
\label{sec:results}
\begin{figure}
    \centering
    \includegraphics[width=0.9\linewidth]{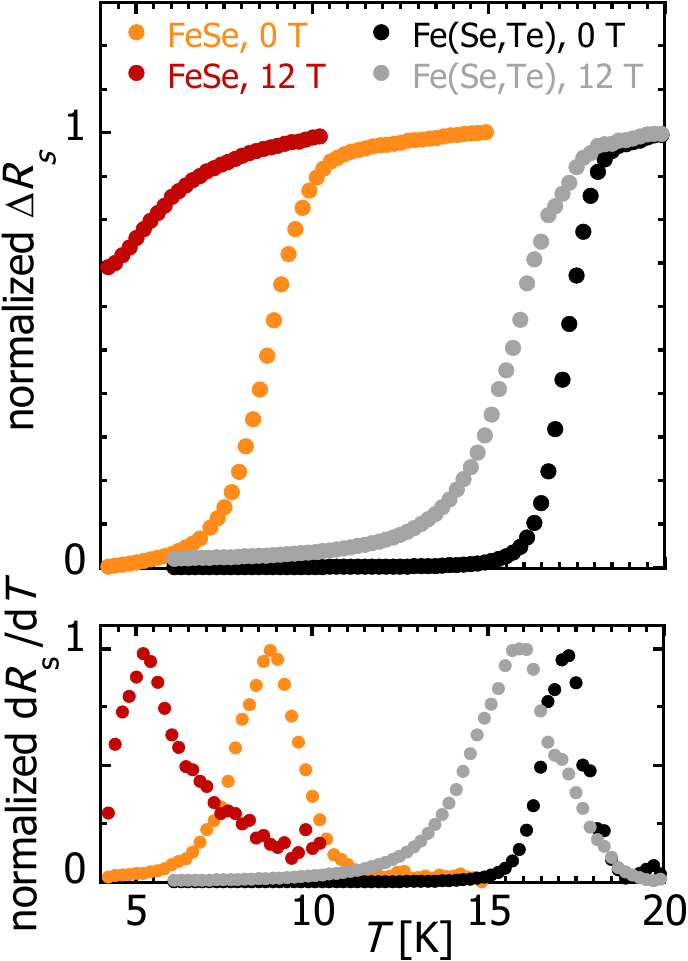}
    \caption{Upper panel: $\Delta R_s(T)$ at the fixed fields $\mu_0H =\qtylist{0;12}{\tesla}$ in FeSe and Fe(Se,Te), normalized with respect to their maximum values $\Delta R_s(T_{max})$. Lower panel: temperature derivative of $R_s$ normalized at the maximum. Symbols are reported in the legend. FeSe has a lower magnetic field resilience with respect to Fe(Se,Te). Data for $dR_s/dT$ have been averaged to reduce noise. Only 5\% of the data are shown to reduce clutter.}
    \label{fig:R_s}
\end{figure}
\begin{figure}
    \centering
    \includegraphics[width=0.9\linewidth]{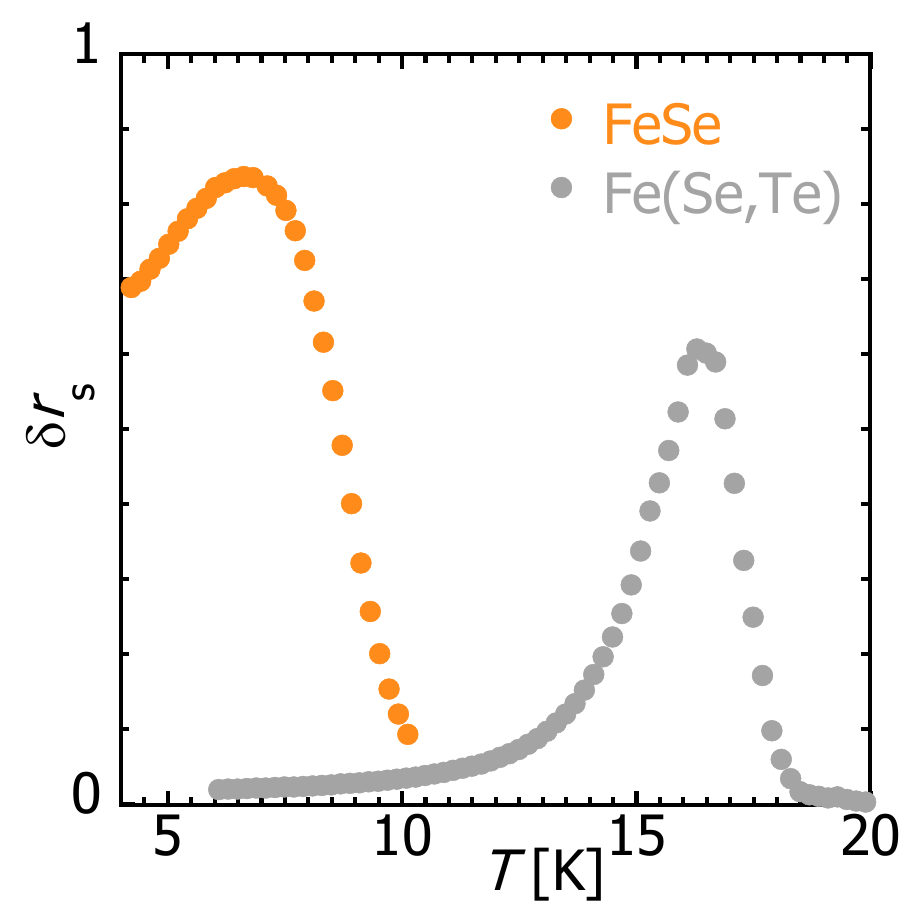}
    \caption{Magnetic-field-induced change in the surface resistance ${\delta r_s=(\Delta R_s(\qty{12}{\tesla})-\Delta R_s(0)})/\Delta R_s(T_{max})$ in FeSe and Fe(Se,Te). Symbols are reported in the legend. Data in FeSe point to a much higher residual level of $\delta r_s$ below the superconducting transition with respect to Fe(Se,Te), an indication of much lower depinning frequency. Only 5\% of the data are shown to reduce clutter.}
    \label{fig:DRH}
\end{figure}

Figure \ref{fig:R_s} reports the main results for what concerns the microwave measurements. There, we report and compare the temperature dependence of $\Delta R_s$ at ${\mu_0H =\qtylist{0;12}{\tesla}}$ in FeSe and Fe(Se,Te). We note that in the normal state $R_s$ in thin films is frequency independent (Equation \eqref{eq:thin}).
FeSe exhibited a marked shift of the superconducting transition with the application of the strong dc magnetic field, with a reduced broadening, much in the same fashion as conventional superconductors. This is a hint for a reduced role of superconducting fluctuations, that are typically dominating in the high part of the transition in cuprates and other high-$T_c$ superconductors. The relatively broad transition even at ${\mu_0H=0}$ is likely  a manifestation of the inhomogeneities in the composition of the sample. The onset of the microwave transition ${T_c\approx \qty{10.5}{\kelvin}}$ reproduces well the onset of the dc transition in a similar film.
Fe(Se,Te) exhibits instead a relatively small shift of the onset of the transition with the magnetic field and a large broadening of the transition (``fan-shaped''), typical of many high-$T_c$ superconductors.  The onset of the microwave transition ${T_c\approx \qty{18}{\kelvin}}$, similar to other samples grown with the same technique \cite{Pompeo_2020_JPCS,Pompeo_2020_Sust,Pompeo_2021_TAS,Magalotti_2025_Arxiv}.
In order to examine in a more quantitative fashion the magnetic resilience of the IBS under study, we calculate, and report in the lower panel of Figure \ref{fig:R_s} the numerical derivative $dR_s/dT$ normalized at the maximum (peak value, $\max(dR_s/dT)$, reached at $T_{peak}$). This quantity is useful to determine the shift of the critical temperature with the field, that can be approximated to the shift of the peak value, and to evaluate the broadening of the transition, as given by the full width at half maximum of the $dR_s/dT$ curves. It is also noticeable that this quantity is able to reveal possible second superconducting phases, that would manifest as ``humps'' or secondary peaks in the curves $dR_s/dT$ vs. $T$ at ${\mu_0H=0}$, and possible significant roles of weak-links, that would manifest as strongly asymmetrical shapes in a magnetic field. 
Then, by examining $dR_s/dT$, we find that with the application of the strong magnetic field ${\mu_0H=\qty{12}{\tesla}}$, the critical temperature shifts by ${\Delta T_c=\qtylist{3.6;1.4}{\kelvin}}$ in FeSe and Fe(Se,Te), respectively, as evaluated by ${\Delta T_c = T_{peak}(\qty{12}{\tesla})-T_{peak}(\qty{0}{\tesla})}$. This is a manifestation of the lower value for the upper critical field in FeSe with respect to Fe(Se,Te). The width $\delta T_c $ of the transition, as evaluated by the full width-half maximum of the curves $dR_s/dT$ vs. $T$, shows a little change in FeSe, ${\delta T_c=\qty{2.1}{\kelvin}}$ irrespective, within the experimental uncertainty, of the application of the magnetic field. In Fe(Se,Te), ${\delta T_c=\qtylist{1.3;2.8}{\kelvin}}$ in zero and \qty{12}{\tesla} magnetic field, respectively. The peculiar behaviour of FeSe require closer inspection, in particular by extending the measurements at a lower temperature, or at lower magnetic field so to explore the field evolution. Nevertheless, this is consistent with the similarity between FeSe and conventional, metallic superconductors. 
A final consideration on the apparent larger effect of the magnetic field on the surface resistance of FeSe. The microwave surface resistance in a dc magnetic field is almost entirely determined by the motion of flux lines. It is useful to calculate the difference ${\delta r_s=(\Delta R_s(\qty{12}{\tesla})-\Delta R_s(0)})/\Delta R_s(T_{max})$, as reported in Figure \ref{fig:DRH}. The quantity $\delta r_s$ represents, at least not too close to $T_c$, the contribution of the vortex motion to the surface resistance, that here coincides with the real part of the complex resistivity, see Equation \eqref{eq:thin}, normalized with respect to the normal state resistivity $\rho_n$. For the real part of the vortex motion complex resistivity, at low enough temperature so that creep effects can be neglected, one writes \cite{gittleman1966radio,pompeo2008reliable}:
\begin{equation}
    \rho_{v1}=\rho_{ff}\frac{1}{1+\left(\frac{\nu_p}{\nu}\right)^2}
    \label{eq:rhov}
\end{equation}
where $\rho_{ff}$ is the flux-flow resistivity and $\nu_p$ is the well-known depinning frequency. For microwave applications in a magnetic field, such as the coating of haloscopes, both a small $\rho_{ff}$ and a large $\nu_p$ are desirable. However, $\rho_{ff}$ depends on intrinsic electronic processes, and it cannot be deliberately engineered, while $\nu_p$ is determined largely by pinning and thus can be somewhat tailored with suitable defects.  Although the precise dependence of $\rho_{ff}$ can be quite complex in IBS \cite{Magalotti_2025_Arxiv}, and a precise determination of $\nu_p$ requires an extensive set of data \cite{Pompeo_2020_Sust}, we can rely on a previous accurate determination of ${\nu_p \sim \qty{60}{\giga\hertz}}$ in Fe(Se,Te) at \qty{6}{\kelvin} and moderate magnetic field \qty{0.6}{\tesla} \cite{Pompeo_2020_Sust}. We observe that at low $T$, $\delta r_s$ is a very small fraction of $\rho_n$ in Fe(Se,Te), while the data suggest that this is no longer the same in FeSe. This is an indication that pinning has ample margins of optimization in FeSe. 
However, it is a first relevant result that FeSe thin films have been successfully grown with satisfactory, albeit not optimized yet, microwave properties especially for what concerns the magnetic field resilience. 
\section{Conclusion}
We have presented a successful growth by PLD of FeSe thin films on dielectric CaF$_2$ substrates. The critical temperature exceeded the reported values for bulk FeSe, most likely because of the induced strain due to the substrate. A sample was selected for measurements of the microwave surface resistance. Measurements performed by the dielectric-loaded resonator method showed a transition in agreement with the dc results. The microwave transition exhibited a relatively large shift with the application of a strong dc magnetic field, but a small broadening, resembling the behavior of conventional metallic superconductor. No evident signatures of weak-link related dissipation were found. The comparison with a thin film of the parent compound Fe(Se,Te) showed that the microwave pinning in FeSe is rather low, and that it must be optimized before application as a coating in haloscopes.
\section*{Acknowledgments}
Support by Federico Quagliata Tamisari for the AFM operation is gratefully acknowledged.
\bibliographystyle{IEEEtran}

\end{document}